%% file: sample.tex
\documentclass[fleqn,10pt]{wlscirep}
\usepackage[utf8]{inputenc}
\usepackage[T1]{fontenc}
\usepackage{bm}
\usepackage{array}
\usepackage{amsmath}
\usepackage{graphicx}
\usepackage{geometry}
\usepackage{xcolor}

\title{Collaborative Expert LLMs Guided Multi-Objective Molecular Optimization}

\author[1,5*]{Jiajun Yu}
\author[2,*]{Yizhen Zheng}
\author[2,*]{Huan Yee Koh}
\author[3,*]{Shirui Pan}
\author[4,5]{Tianyue Wang}
\author[1,+]{Haishuai Wang}

\affil[1]{College of Computer Science and Technology, Zhejiang University, Hangzhou, China}
\affil[2]{Department of Data Science and Artificial Intelligence, Monash University, Victoria, Australia}
\affil[3]{School of Information and Communication Technology, Griffith University, Queensland, Australia}
\affil[4]{Innovation Institute for Artificial Intelligence in Medicine, College of Pharmaceutical Sciences, Zhejiang University, Hangzhou, Zhejiang, China}
\affil[5]{Shanghai Innovation Institute, Shanghai, China}

\affil[*]{These authors contributed equally to this work.}
\affil[+]{Corresponding author: haishuai.wang@zju.edu.cn}

\newcommand{\model}{MultiMol}

\begin{abstract}
Molecular optimization is a crucial yet complex and time-intensive process that often acts as a bottleneck for drug development. Traditional methods rely heavily on trial and error, making multi-objective optimization both time-consuming and resource-intensive. Current AI-based methods have shown limited success in handling multi-objective optimization tasks, hampering their practical utilization. To address this challenge, we present \model, a collaborative large language model (LLM) system designed to guide multi-objective molecular optimization. \model\ comprises two agents, including a data-driven worker agent and a literature-guided research agent. The data-driven worker agent is a large language model being fine-tuned to learn how to generate optimized molecules considering multiple objectives, while the literature-guided research agent is responsible for searching task-related literature to find useful prior knowledge that facilitates identifying the most promising optimized candidates. In evaluations across six multi-objective optimization tasks, \model\ significantly outperforms existing methods, achieving a 82.30\% success rate, in sharp contrast to the 27.50\% success rate of current strongest methods. To further validate its practical impact, we tested \model\ on two real-world challenges. First, we enhanced the selectivity of Xanthine amine congener (XAC), a promiscuous ligand that binds both $A_1R$ and $A_{2A}R$, successfully biasing it towards $A_1R$. Second, we improved the bioavailability of Saquinavir, an HIV-1 protease inhibitor with known bioavailability limitations. Overall, these results indicate that \model\ represents a highly promising approach for multi-objective molecular optimization, holding great potential to accelerate the drug development process and contribute to the advancement of pharmaceutical research. 
\end{abstract}

\begin{document}

\flushbottom
\maketitle

\thispagestyle{empty}

\input{introduction}
\input{main_method}
\input{experiments}

\bibliography{sample}

\input{method}

\input{related}
\input{extented_data}

\end{document}

%% file: introduction.tex
\section*{Main}
Molecular optimization, the process of altering a molecule's structure to enhance properties such as efficacy \cite{lu2010drug}, stability \cite{klepeis2003integrated}, or reduced toxicity \cite{narsinghani2014lead}, is a critical yet challenging stage in drug discovery. This process typically begins after identifying a hit—a molecule with initial activity against a biological target—and progresses through hit-to-lead and eventually to lead optimization, where drug-like properties are fine-tuned to achieve the balance required for clinical success. Despite significant investment, the process often relies on trial-and-error, resulting in suboptimal ligands and contributing to the 90\% failure rate of drug candidates during clinical trials~\cite{sadybekov2023computational}. The inherent complexity of this process lies in optimizing multiple interconnected properties, where even minor structural changes can significantly impact a molecule’s behavior and often lead back to suboptimal outcomes, making molecular optimization a major challenge in drug discovery.

Existing AI methods \cite{liu2024conversational, liu2023multi} have emerged as potential solutions to address these challenges in molecular optimization. However, these methods often underperform in multi-objective optimization and tend to overlook critical constraints, such as validity and scaffold consistency of their generated molecules. As a result, the molecules produced by these models may fail to retain the original scaffold, which is essential for preserving the core biological activity of a compound, or may even be chemically invalid. This mismatch between the capabilities of AI methods and the practical demands of drug discovery highlights a key limitation: optimizing molecular properties while maintaining scaffold consistency is indispensable but remains inadequately addressed. Additionally, the optimization processes of existing methods lack guidance from prior knowledge. current AI optimization approaches often lack guidance from prior knowledge, contrasting sharply with the expert-driven strategies of medicinal chemists, who rely on intuition built from extensive chemical expertise from prior experience and literature-based insights. This expert-driven approach highlights the gap between current AI methods and the practical needs in drug discovery, where guided optimization and scaffold preservation are essential.

Recently, large language models (LLMs) \cite{taylor2022galactica,yang2024qwen2,touvron2023llama,achiam2023gpt} have gained significant attention due to their remarkable capabilities. Trained on vast amounts of data, these models have developed emergent abilities that enable them to perform complex tasks, such as role-playing and reasoning at a level comparable to human experts such as chemists~\cite{mirza2024large} and physicians~\cite{singhal2023large}. In some cases, LLMs have matched or even surpassed human expert performance across a wide range of tasks, such as medical question answering and complex problem-solving. Notably, collaborative LLM agents systems, where multiple agents work together, have demonstrated enhanced performance on intricate tasks like software development~\cite{hong2023metagpt} and healthcare system simulation~\cite{li2024agent}. By leveraging collective knowledge and reasoning skills, these collaborative agents can tackle complex, multifaceted problems more effectively.

Inspired by the advancements in LLMs, we aim to leverage their domain knowledge, information retrieval capabilities and expert-like reasoning to guide molecular optimization more effectively, moving away from random guesswork. To this end, we introduce \model, a collaborative LLM-based system designed to enhance molecular optimization tasks, with a particular focus on multi-objective optimization. \model\  is a dual-agent synergy system consisting of two expert LLM agents: a data-driven worker agent and a literature-guided research agent. The data-driven worker agent is fine-tuned using our specially designed masked-and-recover strategy. This simple yet effective approach alleviates the need for molecular pair editing data, which is commonly required by many molecular optimization models. This fine-tuning equips the worker agent to process an input molecule and desired property specifications, generating diverse, optimized molecular candidates. The literature-guided research agent complements the worker agent by performing targeted web searches to identify relevant molecular characteristics associated with the desired properties. It then aids in filtering the optimized candidates by aligning them with identified molecular characteristics, such as functional groups known to enhance solubility. Optimized molecules possessing these desired characteristics are picked as output.

As a result, \model\ significantly outperforms existing baselines across all molecular optimization tasks. In particular, for multi-objective optimization, while baseline methods achieve a success rate of approximately 10\%, \model\ achieves an average success rate of 66.49\% across all tasks. To highlight its practical utility, we applied \model\ to two real-world scenarios: Saquinavir, an HIV small molecule drug with bioavailability challenges, and Xanthine amine congener (XAC), a non-selective adenosine receptor antagonist requiring improved selectivity. Using \model, we achieved successful optimization in both cases. For Saquinavir, \model\ improved bioavailability while preserving its binding affinity to HIV-1 protease. For XAC, \model\ enhanced its binding affinity to $A_1R$ while dramatically reducing affinity to $A_{2A}R$, thereby improving selectivity. These outcomes underscore the practical value of \model\ and represent a meaningful step from theoretical applications to real-world impact.

%% file: main_method.tex
\section*{Results}
\subsection*{Collaborative LLMs Guided Multi-objective Molecular Optimization}
\begin{figure}
    \centering
    \includegraphics[width=0.85\linewidth]{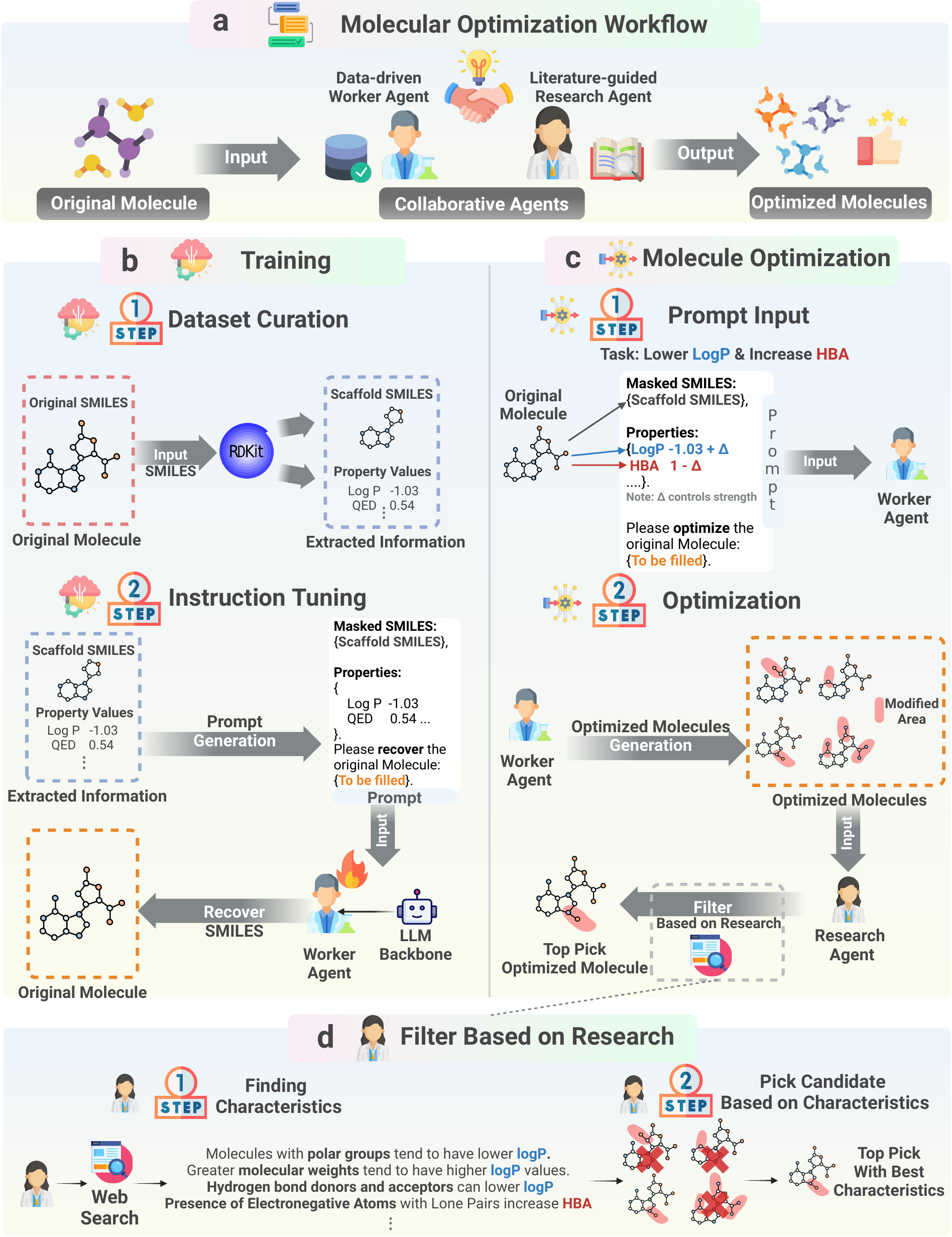}
    \caption{\textbf{\model\ architecture.} (\textbf{a}) \textbf{Molecular Optimization Workflow.} The end-to-end molecular optimization workflow, illustrating how a data-driven worker agent and a literature-guided research agent collaborate to transform an original molecule into a pool of optimized candidates. (\textbf{b}) \textbf{Training.} The training procedure, consisting of two steps: dataset curation (step 1), where RDKit is used to extract scaffold SMILES and property values from the original SMILES, and instruction tuning (step 2), where a large language model is prompted to recover the original molecule based on scaffold SMILES and properties.
(\textbf{c}) \textbf{Molecule Optimization} The multi-objective optimization pipeline, which takes the target molecule's scaffold SMILES and desired properties (step 1), then generates a pool of optimized molecules (step 2). (\textbf{d}) \textbf{Filter Based on Research.} The final research-based filtering stage, where literature-derived insights are used to define key molecular characteristics that guide the selection of the top-performing molecules.}
    \label{fig:model}
\end{figure}

\model\ introduces a framework for learning and executing multi-objective optimization tasks for molecules, addressing the limitations of existing methods that often yield suboptimal results. This system employs two specialized LLM agents: a data-driven worker agent and a literature-guided research agent. The molecule optimization workflow begins by selecting a molecule to be optimized, which the agents collaboratively optimize to enhance desired properties, and output optimized molecules with desired properties (Figure~\ref{fig:model}a).

\noindent \textbf{Training.} Our training procedure comprises two main stages: (1) \textit{pretraining dataset curation} and (2) \textit{instruction tuning} (see Fig. \ref{fig:model}b). During \textit{pretraining dataset curation}, we leverage RDKit~\cite{landrum2013rdkit} to extract both the scaffold (i.e., the core molecular framework) and key molecular properties (e.g., LogP and QED) from each molecule’s SMILES string. From this process, we construct a million size large pretraining dataset where each entry contains three critical pieces of information: the original SMILES string, the corresponding scaffold SMILES string, and the associated property values. Next, we perform \textit{instruction tuning} to fine-tune our data-driven worker agent, which is formed by prompting a LLM backbone (e.g., Galactica-6.7b~\cite{taylor2022galactica} or Llama~\cite{touvron2023llama}) to recover the original SMILES string given its scaffold SMILES string and specified properties. Here, we deliberately opted to use general LLMs, which are trained extensively on diverse textual data—because they can interpret scientific language effectively~\cite{zheng2024large, taylor2022galactica, mirza2024large} while also drawing on broad domain knowledge beneficial for tasks such as molecular optimization. In contrast, specialized models (e.g., those trained only on chemical SMILES) tend to be more narrowly focused and less able to utilize prior knowledge. This instruction tuning approach is particularly advantageous for multi-objective molecule optimization despite its simplicity: the worker agent is explicitly instructed to generate molecules that not only meet given multiple property requirements (e.g., LogP, QED) but also preserve the original molecular scaffold. This ensures that structural integrity is maintained while allowing for optimal modification to improve molecules' properties.


\noindent \textbf{Molecule Optimization.} After training, molecules can be optimized through a two-step process that involves collaboration between a data-driven worker agent and a literature-guided research agent (see Fig. ~\ref{fig:model} (\textbf{b})). In the first step, prompt input, the target molecule’s scaffold SMILES and properties are extracted using RDKit. For instance, if the goal is to reduce LogP while increasing the hydrogen bond acceptor (HBA) count, say the original properties are LogP = -1.03 and HBA = 1, these property values are adjusted using a parameter $\Delta$ that controls the optimization strength (details on $\Delta$ are provided in the Methods section). The scaffold SMILES and the updated property targets are then fed to the worker agent, which is tasked with generating candidate molecules that satisfy the new specifications.

In the second step, optimization, the worker agent generates a pool of optimized molecules, which are subsequently passed to the research agent. The research agent, also created by prompting a large language model backbone, evaluates and selects the best molecules based on literature-derived insights and scientific reasoning. In this work, GPT4o was chosen for its advanced capabilities in information retrieval and domain-specific reasoning. After filtering and selection, the top-performing molecules are identified as the final optimized picks.

\noindent \textbf{Filter Based on Research.} The literature-guided research agent filters the set of candidate molecules generated by the worker agent through a research-informed filtering approach consisting of two steps: identifying key molecular characteristics and selecting candidates based on those characteristics (see Fig.~\ref{fig:model}(\textbf{d})).

In the first step, the research agent uses web search engine (e.g., Google) to gather insights on properties of interest. For example, if the goal is to reduce LogP, the agent investigates known features of low-LogP molecules, such as the presence of polar groups or a relatively high molecular weight. If the aim also includes increasing the hydrogen bond acceptor (HBA) count, the agent searches for characteristics linked to high HBA, such as the presence of electronegative atoms. Based on this information, the agent autonomously constructs a simple linear filtering function. It uses RDKit to check whether candidate molecules exhibit the identified features: each SMILES string is transformed into a feature vector indicating whether those features are present, and the linear function is trained on a small randomly sampled dataset. The detailed feature to estimate each property can be found in Extended Data Table 1. In the second step, the trained function ranks the candidate molecules by predicting scores for each, enabling the research agent to select those most likely to meet the desired property requirements.

%% file: experiments.tex
\subsection*{Performance Evaluation}
We conducted an evaluation of our molecular optimization framework through rigorous experimentation across diverse scenarios, encompassing 6 multi-objective and 8 single-objective optimization tasks, such as increasing the LogP or increasing tPSA and reducing LogP simultaneously. The detailed task captions are shown in Extended Data Table 2. Our systematic analysis revealed \model\ achieved significant improvements across all key performance metrics compared with baselines.The key metrics include hit ratio (loose) and hit ratio (strict), which evaluate the percentage of generated molecules that meet the target properties under loose and strict thresholds, respectively. Notably, the threshold for hit ratio (loose) is uniformly set to 0, whereas the threshold for hit ratio (strict) can be determined by referring to Extended Data Table 2. Additionally, the same scaffold ratio measures the retention of molecular scaffolds, while the valid ratio verifies the validity of the generated molecules. 

In multi-objective optimization tasks (Figure \ref{fig2}a, top), our method successfully optimized several molecular properties at the same time. The radar plot highlights improvements in key metrics: the loose hit ratio increased by 55\%, and the strict hit ratio improved by 54\% compared to the strongest baseline, MoleculeSTM. The same scaffold ratio stayed around 90\%, and the valid ratio was nearly 100\%, indicating strong preservation of molecular structure. This shows that our model not only generates valid molecules with the desired properties but also retains the original molecular scaffold for most cases. Maintaining scaffold consistency is a critical requirement in molecular optimization. This level of performance in generating desired molecules can be attributed to our effective instruction tuning with the worker agent and its synergy with the research agent providing prior knowledge guidance. 
A graphical illustration of the optimization effectiveness of \model\ is shown in Figure \ref{fig2}b. The figure displays scatter plots where each point represents a molecule. The x-axis and y-axis show combinations of properties, including \# HBA-LogP, \# HBD-LogP, and tPSA-LogP. Purple points represent molecules before optimization, while blue points show the optimized ones. Orange arrows indicate the optimization direction of molecules generated by \model\, and red arrows on the axes point to the target optimization directions for the task. We observed that the molecules generated by \model\  aligned well with the multi-objective optimization directions. This demonstrates \model's capability to handle multi-task optimization effectively. 


For single-objective optimization tasks (Figure \ref{fig2}a, bottom), the results show that the hit ratio (loose) reached 90.92\%, surpassing the baseline MoleculeSTM by 43\%, while the hit ratio (strict) achieved 81.96\%, representing a 53\% improvement. The method maintained a valid ratio of nearly 100\% and a same scaffold ratio of approximately 90\%. In single-objective optimization, MultiMol still outperformed the baseline by a large margin, demonstrating that MultiMol delivers stable performance across both single- and multi-objective tasks. Additionally,
MultiMol achieved the same level of valid ratio and scaffold consistency as in multi-objective optimization. The property distribution analysis using density plots (Figure \ref{fig2}c) shows the shift in molecular properties from the initial distribution (purple) to the optimized distribution (blue). Orange arrows indicate the optimization direction of molecules generated by MultiMol, while red arrows on the x-axis represent the expected optimization direction for the task. Overall, the generated molecules align well with the expected optimization direction across all tasks. Notably, in the QED improvement task, the distribution of optimized molecules appears similar to that of the original molecules. This is because the original molecules had an average QED value of 0.76, which is already quite high, leaving limited room for improvement. As a result, other baselines (MoleculeSTM) achieved a maximum accuracy of only 36.52\% for this task, whereas MultiMol achieved an accuracy of 59.69\%. This demonstrates that MultiMol maintains strong performance even on challenging tasks.
We further analyzed the model's ability to recover molecular properties from scaffolds (Extended Data Figure 1). By comparing distributions between source molecules and those reconstructed using only scaffold information and property targets, we found remarkably consistent patterns. The density plots reveal highly similar distributions across all key properties (MolLogP, QED, TPSA, NumHAcceptors, NumHDonors), demonstrating that the model effectively captures and preserves property distributions during reconstruction. This ability to faithfully reconstruct molecules while maintaining their desired characteristics suggests the model has developed a deep understanding of meaningful molecular distributions, which contributes to its strong performance in multi-objective optimization tasks.

Besides, to demonstrate that the effective performance of MultiMol is not due to the LLM having seen these molecules before, we show in Extended Data Table 3 that the results on a batch of newly released molecules (October 24, 2024) are similar to those on the current validation set, indicating that the superior performance of MultiMol is not due to data leakage.

\subsection*{Ablation study}
\subsubsection*{Effect of the Filter}
Figure \ref{fig2}.d illustrates the efficacy of our Research Agent's filtering mechanism through a comparative analysis of property changes ($\Delta$ Value) between the initial candidate pool (purple) and final selected top pick molecules (blue). The results demonstrate that our filtering approach effectively identifies optimized molecules across multiple objectives. 
For example, in the optimization task 201, while unfiltered candidates exhibited an average LogP reduction of -1.51 units, our filtered candidates achieved a more substantial -2.52 unit reduction, surpassing the average by 1.01 units. This improvement showcases the filtering mechanism's effectiveness to identify candidates with desired target properties.

However, for the task 205, which involved simultaneous optimization of tPSA and LogP, our selected candidates showed a marginally higher $\Delta$ LogP (0.11 units above average) compared to the general candidate pool. This apparent discrepancy reflects an intentional trade-off, as tPSA and LogP typically exhibit an inverse relationship, as documented by Matsson and Kihlberg \cite{matsson2017big}. In this case, our filtering mechanism strategically prioritized candidates with lower $\Delta$ tPSA values, recognizing that tPSA reduction presents a greater optimization challenge than LogP modification for this particular task.

Additionally, we analyzed the impact of candidate pool size on filtering effectiveness (Extended Data Figure 2). As the pool size increased from 1 to 8 candidates, optimization performance improved substantially, with hit ratios showing marked increases. Importantly, both valid ratio and same scaffold ratio remained consistently high (~98\% and ~90\% respectively) across all pool sizes, demonstrating MultiMol's ability to reliably generate valid molecules while preserving scaffold integrity regardless of the number of candidates generated.

\subsubsection*{Effect of the Backbone}
Figure \ref{fig2}e and Extended Data Figure 3 present performance comparisons across different backbone architectures after fine-tuning with 1 million molecules (which is randomly sampled from PubChem \footnote{https://pubchem.ncbi.nlm.nih.gov}). Using consistent training data across all experiments, we observed significant performance improvements as the model scale increased, aligning with the scaling laws observed in the LLM domain \cite{chowdhery2023palm}. The 125M model achieved modest performance with a hit ratio (loose) of 12.01\% and hit ratio (strict) of 3.70\%, while maintaining a valid ratio of 92.25\% and same scaffold ratio of 60.52\%. The 1.3B model showed improved metrics with a hit ratio (loose) of 31.26\% and hit ratio (strict) of 16.1\%. Most notably, the 6.7B parameter model substantially outperformed smaller models, achieving hit ratio (loose) of 73.62\% and hit ratio (strict) of 57.37\%, while maintaining excellent valid ratio (98.33\%) and same scaffold ratio (95.40\%). Further scaling to 30B parameters yielded comparable results (hit ratio (loose): 73.78\%, hit ratio (strict): 56.31\%, valid ratio: 97.67\%, same scaffold ratio: 96.14\%), suggesting diminishing returns beyond 6.7B parameters.

Additionally, we compared three mainstream language models of similar size: Qwen2.5-7b, Llama3.1-8b, and Galactica-6.7b, a model specifically pre-trained on scientific literature. As shown in Extended Data Figure 3, Galactica significantly outperformed the other two backbones across all four metrics. This finding suggests that stronger general language capabilities do not necessarily translate to better performance in specific domains. Rather, language models specifically pre-trained on life sciences literature, like Galactica, tend to achieve superior results in domain-specific tasks like molecular optimization.

\subsubsection*{The Scale of the pretraining Dataset}
\begin{figure}
    \centering
    \includegraphics[width=1.0\linewidth]{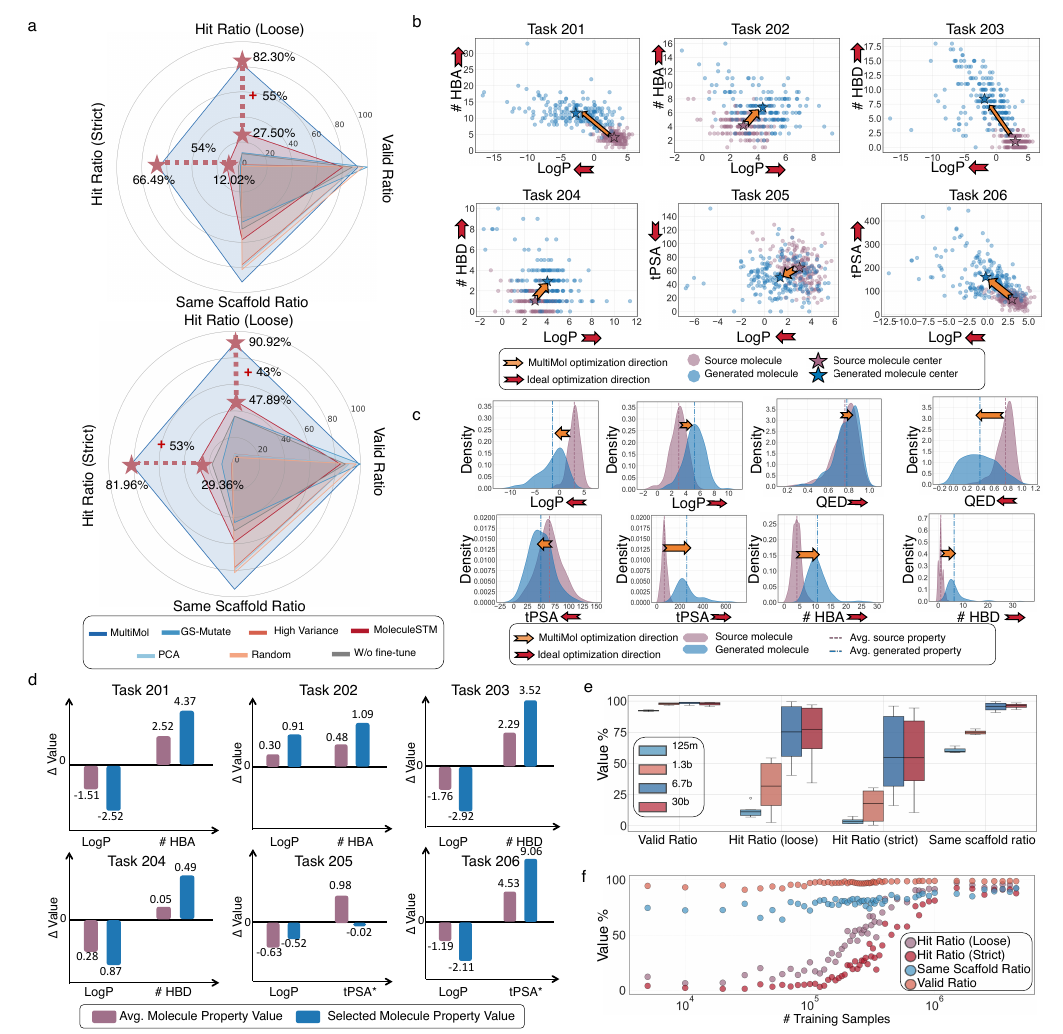}
    \caption{\textbf{\model\ performance.} 
(\textbf{a}) \textbf{Baseline Comparison.} Radar plots comparing \model\ with baseline models across four key metrics for multi-objective (top) single-objective (bottom) optimization tasks, where higher values indicate better performance. 
(\textbf{b}) \textbf{Mult-objective Optimization Visualization.} Scatter plot visualization of multi-objective optimization results on the test set, showing initial molecules (purple dots) and optimized molecules (blue dots) in property space. Stars indicate the centroids of initial and optimized populations, with red arrows denoting the desired optimization directions.
(\textbf{c}) \textbf{Single-objective Optimization Visualization.} Property distribution plots for single-objective optimization on the test set, with property values on x-axis and density on y-axis. Purple and blue curves represent distributions before and after optimization, respectively.
(\textbf{d}) \textbf{Comparison of Filtered vs. Unfiltered Molecules.} Comparison of average property changes between unfiltered generated molecules (purple) and Research Agent filtered candidates (blue), demonstrating the effectiveness of our filtering approach in molecular optimization. Property values are normalized (tPSA* = tPSA/10) for visualization clarity.
(\textbf{e}) \textbf{Backbone Size.} Performance comparison across different backbone architectures after fine-tuning.
(\textbf{f}) \textbf{Dataset Scale} Effect of training dataset size on model performance, evaluated using multiple metrics on the test set.
}
    \label{fig2}
\end{figure}

The relationship between pretraining dataset size (measured by the number of training samples) and model performance after fine-tuning is illustrated in Figure \ref{fig2}f. Using our largest 6.7B parameter model as the backbone architecture, we observed that valid ratio and same scaffold ratio remained relatively stable across different training dataset sizes, ranging from 93-99\% and 72-88\% respectively. However, the hit ratios showed significant improvements with increased training data. With 10,000 training samples, the model achieved modest performance with hit ratio (loose) of 6.44\% and hit ratio (strict) of 2.15\%. Increasing to 100,000 training samples improved these metrics substantially, with hit ratio (loose) reaching 25.35\% and hit ratio (strict) rising to 5.66\%. Most notably, scaling to 1 million training samples yielded dramatic improvements in hit ratios (loose) increased to 91.54\% and hit ratio (strict) to 81.19\%. The hit ratio metrics plateaued beyond 1 million samples, suggesting this training dataset size achieves an optimal balance between computational efficiency and model performance while maintaining consistently high valid ratio and sames scaffold ratio throughout.

\subsection*{Molecular Optimization Case Examples}

\begin{figure}
    \centering
    \includegraphics[width=1.0\linewidth]{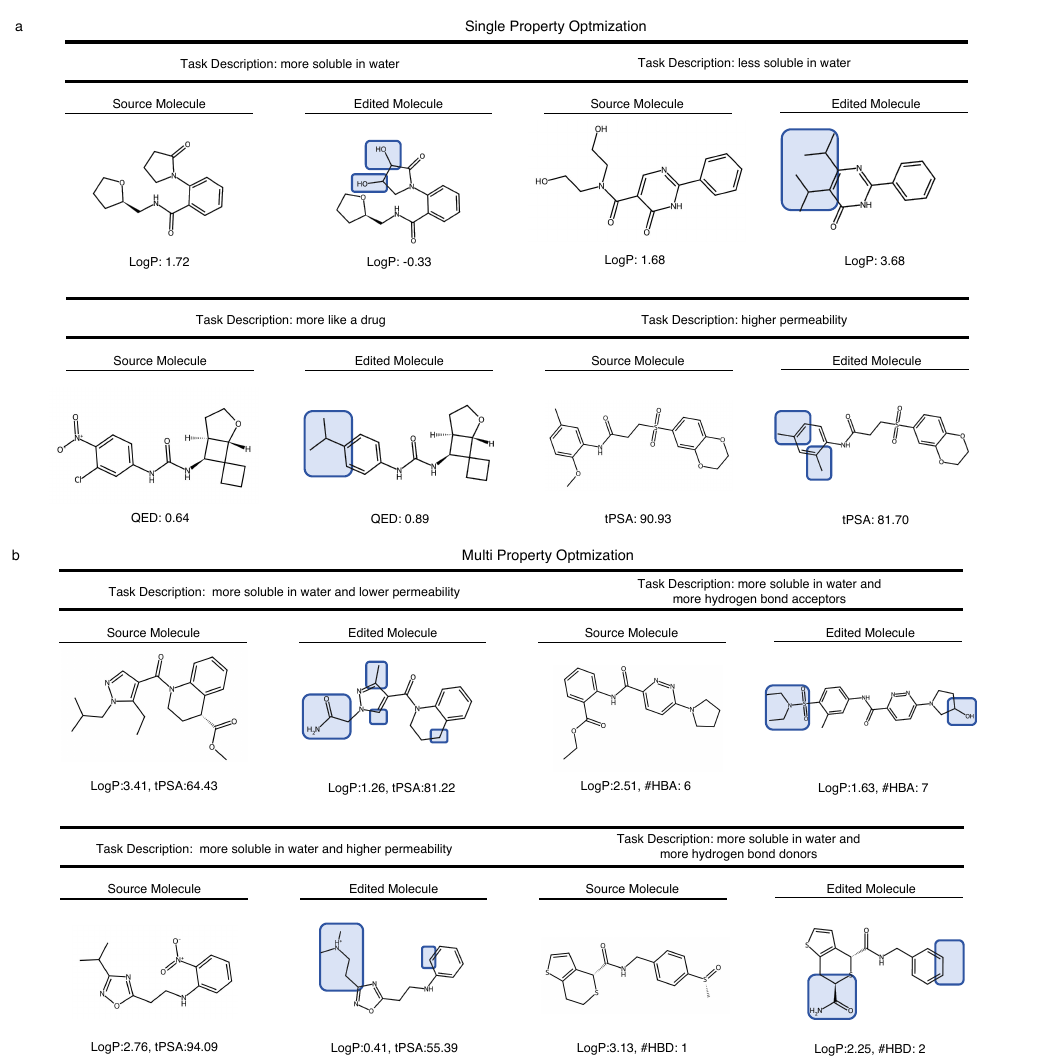}
    \caption{\textbf{Representative examples of molecular optimization using \model.} The figure showcases paired source and optimized molecules, demonstrating both single-property (\textbf{a}) and multi-property (\textbf{b}) optimization tasks. For each pair, source molecules (left) display initial property values, while optimized molecules (right) show improved values after modification. Blue highlights emphasize key structural changes implemented by \model\ to achieve the desired property improvements while maintaining molecular scaffolds.}
    \label{figure6}
\end{figure}

Figure \ref{figure6} illustrates representative molecular optimization outcomes achieved by \model. The figure presents paired source and optimized molecules, with source molecules (left) and their corresponding optimized structures (right) accompanied by quantitative property values. Structural modifications are highlighted in blue to facilitate comparative analysis.
Figure \ref{figure6}.a illustrates successful cases of single-property optimization. In solubility optimization experiments, the systematic incorporation of hydrophilic moieties, such as the introduction of hydroxyl groups (highlighted in blue), led to measurable reductions in LogP values, consistent with the existing knowledge \cite{breslow1991hydrophobic}. For drug-likeness optimization (QED), the model effectively removed synthetically challenging functional groups while preserving the core molecular scaffold and essential pharmacological features, thereby maintaining structural integrity.

Figure \ref{figure6}.b exhibits the model's efficacy in multi-parameter optimization, specifically addressing simultaneous modifications of LogP, hydrogen bonding properties (hydrogen bond acceptors/donors), and topological polar surface area (tPSA). The implemented structural modifications achieved targeted property improvements while preserving scaffold integrity. 
In one illustrative case, we successfully reduced LogP from 2.51 to 1.63 while simultaneously increasing the number of hydrogen bond acceptors from 6 to 7 through strategic incorporation of nitrogen-containing heterocyclic rings. Another representative example demonstrates concurrent optimization of solubility and hydrogen bonding capacity, where LogP was decreased from 3.13 to 2.25 and hydrogen bond donors were increased from 1 to 2 through targeted addition of -NH and -OH functional groups at key positions.


Besides, Extended Data Figure 4 illustrates the changes in functional group distributions during molecular optimization across different tasks. For Task 101 (decreasing LogP), hydrophilic groups (-OH, -NH\textsubscript{2}, -COOH) show positive $\Delta$Scores while hydrophobic groups (-CH\textsubscript{3}, aromatic rings) show negative $\Delta$Scores, aligning with established principles for improving water solubility. In Task 105 (decreasing tPSA), the model systematically reduced polar functional groups while maintaining non-polar groups. Tasks 107 and 108 demonstrate selective increases in hydrogen bond acceptors (N, O-containing groups) and donors (-OH, -NH groups) respectively. These systematic modifications across tasks validate that our model effectively applies fundamental chemical principles during optimization.

\subsection*{Optimization in Real-world Drug Discovery Scenarios} 
Having established the effectiveness of MultiMol, we applied it to optimize drug candidates with suboptimal ligand properties in real-world scenarios: Xanthine Amine Congener (XAC), a promiscuous ligand requiring improved receptor selectivity, and Saquinavir, an FDA-approved drug with limited solubility and suboptimal binding affinity.

\begin{figure}
    \centering
    \includegraphics[width=1.0\linewidth]{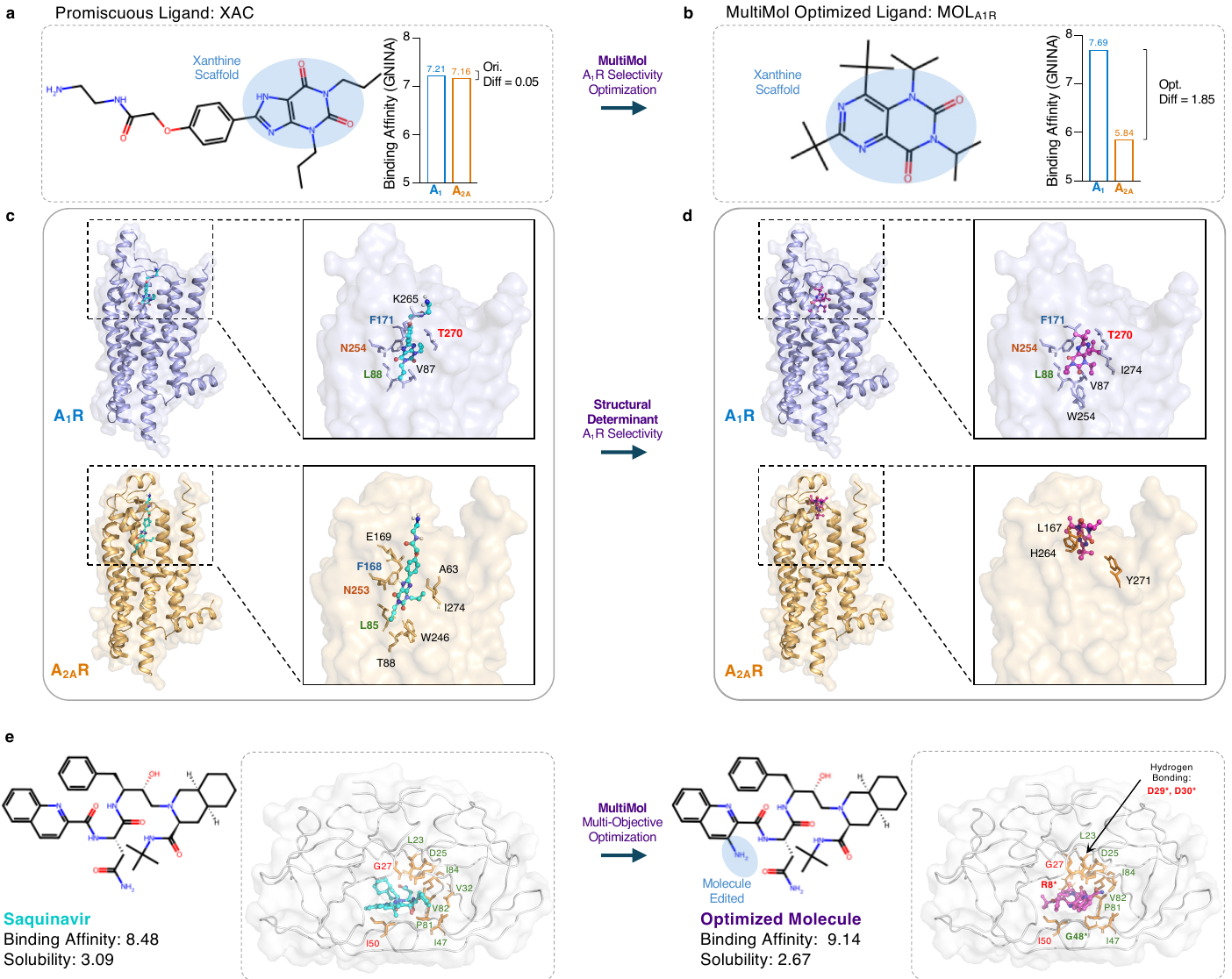}
    \caption{
    \textbf{\model-guided optimization of adenosine receptor subtype selectivity.} (\textbf{a}) The promiscuous ligand XAC (cyan) binds both A\textsubscript{1}R (light blue) and A\textsubscript{2A}R (light orange) with minimal selectivity with a binding affinity difference of $\Delta G = 0.05$ kcal/mol as determined by GNINA. XAC’s xanthine scaffold and interactions with conserved residues, including N254 in A\textsubscript{1}R and N253 in A\textsubscript{2A}R, contribute to its lack of receptor discrimination. (\textbf{b}) MultiMol-guided optimization of XAC produced MOL\textsubscript{A1R} (purple-red), a selective ligand with a significant increase in binding affinity difference favoring A\textsubscript{1}R, GNINA $\Delta G = 1.85$ kcal/mol. Structural modifications targeted A\textsubscript{1}R-specific features to enhance selectivity. (\textbf{c}) XAC binds A\textsubscript{1}R and A\textsubscript{2A}R by engaging conserved residues, including $\pi$-stacking with F171 in A\textsubscript{1}R and F168 in A\textsubscript{2A}R, and hydrogen bonding with N254 in A\textsubscript{1}R and N253 in A\textsubscript{2A}R. However, promiscuity arises due to limited engagement with A\textsubscript{1}R-specific residues such as T270. (\textbf{d}) The optimized ligand MOL\textsubscript{A1R} exploits A\textsubscript{1}R-specific structural features, forming enhanced hydrophobic interactions with residues including T270, W247, and I274, while avoiding favorable interactions with A\textsubscript{2A}R-specific residues such as H264 and Y271. (\textbf{e}) Optimization of Saquinavir showing the original molecule structure and binding mode (left) with binding affinity of 8.48, and the optimized molecule (right) with improved binding affinity of 9.14 and enhanced solubility profile.}
    \label{fig:case_study}
\end{figure}

\subsubsection*{XAC Selectivity Optimization}
In drug discovery, small-molecule ligands often exhibit promiscuity by binding to multiple targets, potentially compromising therapeutic specificity and increasing off-target effects~\cite{roth2004magic,hopkins2009predicting}. To address this challenge, we focused on optimizing XAC, a well-characterized small molecule that binds to both A\textsubscript{1}R and A\textsubscript{2A}R, to enhance its selectivity for A\textsubscript{1}R. This optimization is particularly important for adenosine A\textsubscript{1} receptor (A\textsubscript{1}R), a G-protein-coupled receptor (GPCR) crucial for cardiovascular and neurological functions~\cite{draper2018structure}, where achieving subtype selectivity is difficult due to the high conservation of the orthosteric binding site across adenosine receptor (AR) subtypes: A\textsubscript{1}R and A\textsubscript{2A}R.

XAC demonstrates promiscuity by binding to both A\textsubscript{1}R and A\textsubscript{2A}R (Figure \ref{fig:case_study}a). As shown in Figure \ref{fig:case_study}.c, structural analysis revealed that XAC interacts with conserved residues in the orthosteric binding site of both receptors, such as hydrogen bonding with N254 in A\textsubscript{1}R and N253 in A\textsubscript{2A}R, located in similar positions. Additionally, $\pi$-stacking interactions with F171 in A\textsubscript{1}R and F168 in A\textsubscript{2A}R stabilize the ligand’s positioning in the binding pocket, while hydrophobic contacts with residues like W247 and L88 in A\textsubscript{1}R, and W246 in A\textsubscript{2A}R, further contribute to its binding. Despite these shared interactions, the binding pockets of A\textsubscript{1}R and A\textsubscript{2A}R differ significantly in their topology. A\textsubscript{1}R features a wider binding pocket shaped by T270, a residue that acts as a ``gatekeeper'' by accommodating bulkier substituents. In contrast, A\textsubscript{2A}R has a narrower cavity due to the steric hindrance imposed by M270, which limits ligand accessibility to hydrophobic subregions. These structural differences provide a clear opportunity for selectively optimizing ligands toward A\textsubscript{1}R.

To enhance the selectivity of XAC for A\textsubscript{1}R, we utilized MultiMol to optimize XAC by designing a protein-ligand prompt, as detailed in Extended Data Table 4. Optimization targeted key determinants of subtype selectivity by exploiting the wider A\textsubscript{1}R pocket while reducing compatibility with the narrower A\textsubscript{2A}R pocket. Specifically, MultiMol introduced bulky substituents in XAC to target the hydrophobic pocket near T270 in A\textsubscript{1}R, a region inaccessible in A\textsubscript{2A}R due to the steric effects of M270. As shown in Figure \ref{fig:case_study}.d, this modification maximized hydrophobic interactions with residues such as W247 and I274, enhancing the ligand’s binding affinity to A\textsubscript{1}R. Simultaneously, the optimized structure avoided favorable interactions with residues specific to A\textsubscript{2A}R, such as H264 and I274, thereby reducing promiscuity.

Post-optimization structural analysis using the Protein-Ligand Interaction Profiler (PLIP) confirmed that the modified XAC demonstrated enhanced hydrophobic interactions with A\textsubscript{1}R-specific residues, such as T270 and I274, while retaining critical $\pi$-stacking interactions with F171 and hydrogen bonding with N254. The formation of a hydrogen bond with T270 is particularly significant, as it stabilizes the ligand within the unique hydrophobic pocket of A\textsubscript{1}R, further enhancing its affinity and selectivity~\cite{glukhova2017structure}. This interaction likely prevents the ligand from favorably binding to A\textsubscript{2A}R, where the corresponding residue (M270) cannot form such interactions due to steric hindrance. These results validate the hypothesis that MultiMol can achieve subtype selectivity by leveraging data-driven learning and prior knowledge to guide optimization, exploiting spatial differences in receptor binding pockets and selectively targeting receptor-specific residues.

\subsubsection*{Saquinavir Solubility Optimization}
We further validated MultiMol's optimization capabilities using Saquinavir \cite{collier1996treatment}, an FDA-approved HIV protease inhibitor from 1995. Despite its therapeutic potential, Saquinavir's clinical efficacy has been limited by poor bioavailability ($\sim$5\%) \cite{vyas2008improved} due to low water solubility. Using MultiMol, we aimed to enhance both solubility and binding affinity (Figure \ref{fig:case_study}e). With a suble addition of primary amine (-NH\textsubscript{2}) to the original ligand, MultiMol's optimization resulted in improved binding affinity (from 8.48 to 9.14) and enhanced solubility profile (LogP decreased from 3.09 to 2.67). This case study demonstrates MultiMol's versatility in executing both subtle and substantial molecular modifications while preserving the core therapeutic properties. The successful optimization of an established drug like Saquinavir highlights the framework's potential in addressing real-world pharmaceutical challenges through targeted property improvements.

These case studies underscore the successful application of MultiMol in addressing real-world drug discovery challenges. In the first case, MultiMol effectively optimized a promiscuous ligand by transforming XAC into a selective molecule for A\textsubscript{1}R through targeting receptor-specific features while preserving key interactions. In the second case, it improved both the binding affinity and solubility profile of Saquinavir, demonstrating its versatility in handling established drugs. These results highlight the framework's potential in designing selective ligands for complex receptor systems and optimizing drug properties to enhance therapeutic efficacy, showcasing its capacity to advance precision medicine through targeted molecular optimization.

\subsection*{Discussion}
This study introduces MultiMol, a framework that uses collaborative large language models (LLMs) for molecular optimization in drug discovery. Our tests show that MultiMol performs well in both single-property and multi-property optimization tasks. The framework effectively handles molecular structures and achieves good results across different optimization tasks, such as drug-like property enhancement and target selectivity improvement.

MultiMol has shown good performance in real applications, e.g., in optimizing adenosine receptor selectivity. The framework can modify molecular structures while keeping important parts intact and improving target properties. This success is largely due to our carefully designed prompt-tuning paradigm, which allows for flexible modifications. By simply adjusting the properties specified in the prompt, MultiMol can tackle a wide range of tasks with ease. The tasks we tested represent only a small subset of its potential applications. We believe this paradigm can address more complex and diverse challenges in the future.

Though achieving promising results, our framework faces several inherent limitations and challenges. 
First, LLMs are fundamentally constrained by the distribution of their training data, which makes them less effective at generating molecular structures that significantly deviate from their training examples. This data dependency limits their ability to explore truly novel chemical space. We believe this limitation could be addressed by leveraging future LLMs trained on more advanced architectures and significantly larger, more diverse datasets.
Second, the current approach may struggle with complex molecular scaffolds that require a deeper chemical understanding beyond the statistical patterns learned during training. In this regard, integrating additional specialized agents into the framework could help tackle such complex tasks more effectively.
Additionally, while LLMs excel at recognizing patterns and correlations, they often lack the mechanistic understanding of chemical principles required for groundbreaking molecular innovation. We believe that human expertise and innovative thinking remain indispensable in this domain. Collaborative workflows that integrate human creativity with model-driven insights are likely to become a mainstream approach in the future.

The agent-based collaborative approach used in MultiMol shows promise for future molecular design work, as we can add more specialized agents to handle different parts of the optimization process. This approach could be useful in other areas of drug discovery and materials science that need complex optimization. Looking ahead, the field of AI-driven molecular design needs continued development, where combining LLMs with other computational tools and expert knowledge will help advance molecular innovation and lead to new discoveries in drug development and related areas.

%% file: method.tex
\section*{Methods}

\subsection*{Dataset}
For base property optimization tasks, we curated a diverse dataset of one million molecules randomly sampled from PubChem~\cite{kim2021pubchem}. For each molecule, we calculated key physicochemical properties including LogP (lipophilicity), QED (drug-likeness), tPSA (topological polar surface area), number of hydrogen bond acceptors (\# HBA), and number of hydrogen bond donors (\# HBD) using RDKIT~\cite{landrum2013rdkit}. This comprehensive dataset enabled effective pre-training of our model. For evaluation, we utilized the test set from MoleculeSTM~\cite{liu2023multi}, consisting of 200 carefully selected molecules that represent a broad range of chemical space.

For binding affinity optimization tasks, we leveraged the Protein Data Bank (PDB)~\cite{berman2000protein} dataset, which contains experimentally determined three-dimensional structures of protein-ligand complexes along with their binding affinity data.

\subsection*{Baseline}

To rigorously evaluate our approach, we conducted comprehensive comparisons against six state-of-the-art baseline models. Three of these baselines, derived from \cite{liu2022graphcg}, operate in latent representation space before reconstructing molecules:

(i) Random perturbation: Introduces stochastic noise into molecular representations to explore the chemical space
(ii) Principal Component Analysis (PCA): Leverages eigenvector-based latent directions extracted from molecular representations to guide optimization
(iii) High Variance Direction: Exploits the most variable latent dimension through one-hot encoding for targeted molecular editing

We further benchmark against Genetic Search (GS), which performs direct molecular modifications in chemical space. While GS implements the graph genetic algorithm framework \cite{jensen2019graph}, it relies on unguided random search due to the absence of a retrieval database in zero-shot scenarios.
Our evaluation also includes MoleculeSTM \cite{liu2023multi}, a text-prompt driven molecular editing system, alongside an untuned Galactica 6.7B model.

\subsection*{Detailed Architecture of MultiMol}
 
\model\ employs a dual-agent architecture comprising a worker agent and a research agent that synergistically leverage large language models (LLMs) for molecular optimization. The worker agent constructs a comprehensive self-supervised training dataset from PubChem\footnote{https://pubchem.ncbi.nlm.nih.gov/}, encompassing over one million molecules. For each molecule, the agent extracts its scaffold and computes relevant properties via RDKIT, embedding these components into structured prompts following a standardized template. This prompt-based methodology facilitates efficient dataset construction while maintaining critical molecular information integrity.
The system's training objective focuses on reconstructing source SMILES strings from scaffolds and property values, enabling the model to learn intricate structure-property relationships. This approach circumvents the need for expensive molecule pair generation while preserving key substructural elements. During inference, the model generates candidate molecules by incorporating desired properties and extracted scaffolds, maintaining structural integrity throughout the optimization process.

The research agent enhances optimization by providing literature-derived insights and evaluating candidates through empirically-based scoring functions, ensuring generated molecules align with optimization objectives. The following sections detail the specific mechanisms and functionalities of these complementary agents.

\subsubsection*{Worker Agent}
The Worker Agent enhances the LLM's capability to capture molecular distributions through fine-tuning, enabling generation of diverse, valid small molecules that satisfy specified optimization objectives. The agent operates through two primary components: \textit{instruction dataset construction} and \textit{training and inference}.

\noindent \textbf{Instruction dataset construction.} We propose a systematic methodology for constructing a comprehensive self-supervised instruction dataset that facilitates both multi-objective and single-objective molecular optimization through an innovative recover-scaffold paradigm.

Let $m$ denote a molecule with its corresponding molecular scaffold $s^m$, and let $\mathbf{p} = \{p_1, \ldots, p_n\}$ denote the set of molecular properties under consideration. For each property $p_i \in \mathbf{p}$, we compute its corresponding value $v_i^m \in \mathbf{R}$, yielding a property value vector $\mathbf{v}^m \in \mathbb{R}^{|\mathbf{p}|}$ utilizing RDKIT and complementary computational tools. These properties are structured as a source property-value pair string $\text{SourP}$, formally expressed as a text ``$|p_1| v_1^m, \ldots, |p_n| v_n^m$''. Subsequently, we construct instruction prompts adhering to a standardized template: \textit{``Recover Scaffold $[s^m]$ with the $[\text{SourP}]$, where the completed molecule is $[m]$''}. $\text{SourP}$ encapsulates diverse molecular characteristics, thereby enabling the concurrent acquisition of both single-task and multi-task optimization capabilities through a unified learning framework.

We curated a dataset of 1 million small molecules from PubChem, processing each to extract its scaffold and calculate relevant properties. This approach enables generation of diverse prompts through varied property combinations and optimization objectives, while offering two key advantages:
\begin{itemize}
    \item Elimination of computational overhead associated with explicit molecule pair construction through self-supervision
    \item Enhanced preservation of molecular scaffolds during optimization, maintaining essential structural characteristics
\end{itemize}

\noindent \textbf{Training and Inference.} During training, we leverage our efficient prompt design to update the LLM's parameters. Given varying model scales, we employ different parameter-efficient fine-tuning approaches, including QLoRA\cite{}, LoRA\cite{}, and full-parameter fine-tuning. Our concise prompts significantly reduce computational requirements while maintaining performance.

During inference, we employ a systematic approach to molecular generation by modulating property values within prompts. For each property $i$ with initial value $v_{i}^m \in \mathbb{R}$, we construct a target property-value pair string $\text{TarP}$ by incorporating a desired property change $\Delta_i \in \mathbb{R}$ for property $i$, yielding a target value of $v_{i}^m + \Delta_{i}$. This methodology leverages both the model's pre-trained chemical knowledge and fine-tuned property distributions to generate molecules satisfying multiple optimization criteria simultaneously.

To illustrate this process, consider a molecule $m$ with scaffold $s^m$, initial property vector $\mathbf{v}^m$, and property set $\mathbf{p}$. During training, the model receives prompts of the form:

\textit{``Recover Scaffold $[s^m]$ with the property $[p_1]$ of $[v_{1}^m]$, \ldots, the property $[p_n]$ of $[v_n^m]$, the completed molecule is $[m]$''}

For inference, we transform this prompt structure to incorporate the desired property changes:

\textit{``Recover Scaffold $[s^m]$ with the property $[p_1]$ of $[v_{1}^m + \Delta_{1}]$, \ldots, the property $[p_n]$ of $[v_n^m + \Delta_{n}]$, the completed molecule is ?''}
And the detailed $\Delta$ for each task is shown in Extended Data Table 5.

\subsubsection*{Research Agent}

The Research Agent plays a crucial role in guiding the Worker Agent by providing insights from scientific literature, leveraging two primary modules: \textit{Literature Search} and \textit{Guide Worker Agent}. These modules are designed to harness prior knowledge to enhance the molecular optimization process effectively.

\noindent \textbf{Literature Search.} Let $\mathcal{L}$ denote the set of relevant scientific literature retrieved from academic search engines. For each optimization task with target properties $\mathbf{p}$, the Research Agent queries these platforms to obtain a subset of papers $L_{\mathbf{p}} \subset \mathcal{L}$ containing relevant molecular insights. This literature-derived knowledge focuses on identifying structural features that characterize molecules with desired property values.

For each property $p_i \in \mathbf{p}$, the agent analyzes $L_{\mathbf{p}}$ to extract a set of mapping functions $\mathbf{F}_i = \{F_{i1}, F_{i2}, \ldots, F_{ik_i}\}$, where $k_i$ is the number of mapping functions extracted for property $i$ from the literature, and each function takes a molecule's SMILES string as input and outputs a value in $\mathbb{R}$, strongly associated with the property $p_i$. 

\noindent \textbf{Guide Worker Agent.} 
The Research Agent leverages the extracted knowledge to enhance molecular optimization through literature-guided candidate evaluation. Given the identified structural features $\mathbf{F}_i$ for each property $p_i$, we define a systematic scoring framework:

We define a scoring function $F: \mathcal{S} \rightarrow \mathbb{R}$ that approximates the presence of a feature in a molecule, where $\mathcal{S}$ represents the space of valid SMILES strings. This function leverages computed molecular descriptors to estimate specific structural characteristics that are difficult to measure directly. For instance, given a molecule $m'$ generated based on $m$ by \model, the scoring function $F$ calculates the feature for $m'$, and obtains the feature $x^{m'}_i \in \mathbb{R}^{k_i}$ for property $i$ and molecule $m'$. 
Subsequently, we employ a regression-based approach to determine the optimal weights for each feature $x^{m'}_i \in \mathbb{R}^{k_i}$, aiming to accurately predict the target property value. The regression model is trained on the input features $x^{m'}_i$ with the corresponding target property values as the response variable. This process yields the regression function:

\begin{equation}
    G(m') = \sum_{j=1}^{k_i} w_j F_{ij}(m'),
\end{equation}

where $w_j$ represents the learned weight for the $j^{th}$ feature $F_{ij}(m')$.

In the training phase, we generate a dataset of 1000 molecules with known target property values for each property $p_i$. This dataset is used to train a linear regression model, and the learned weights $w_j$ are stored. The training process involves minimizing the mean squared error between the predicted and actual target property values to optimize the weights.

During inference, given a generated molecule $m'$ and the pre-trained linear regression model's weights $w_j$, we compute a score using the regression function $G(m')$. The molecules are then ranked based on their scores, and the top-scoring molecules are selected as the most promising candidates.

\subsubsection*{MultiMol Workflow Example}

To better understand the functionality of MultiMol, we provide a detailed example of its workflow.

\paragraph{Task Objective}
We start with the following molecule as an example:
\begin{itemize}
    \item \textbf{Source Molecule:} \texttt{Cn1ccc(C(=O)Nc2sc3c(c2C\#N)CCC3)cc1=O}
    \item \textbf{Optimization Goals:} Increase water solubility and hydrogen bond acceptors.
    \item \textbf{Initial Properties:} LogP = 2.05, \# HBA = 5.0
\end{itemize}

\paragraph{Workflow Steps}

\begin{itemize}
    
\item Step 1: Scaffold extraction
Using RDKIT, the scaffold of the source molecule was extracted:
\[
\text{Scaffold: } \texttt{O=C(Nc1cc2c(s1)CCC2)c1cc[nH]c(=O)c1}
\]

\item Step 2: Designing prompts for work agent
The following prompts were designed and input into the Work Agent to generate molecules based on the scaffold while optimizing for lower LogP values and higher hydrogen bond acceptor counts:

\begin{verbatim}
<s> Masked molecule [START_SMILES]O=C(Nc1cc2c(s1)CCC2)c1cc[nH]c(=O)c1[END_SMILES],
property: |LogP|1.05, |NumHAcceptors|6. Completed molecule: [START_SMILES]

<s> Masked molecule [START_SMILES]O=C(Nc1cc2c(s1)CCC2)c1cc[nH]c(=O)c1[END_SMILES],
property: |LogP|0.05, |NumHAcceptors|7. Completed molecule: [START_SMILES]

<s> Masked molecule [START_SMILES]O=C(Nc1cc2c(s1)CCC2)c1cc[nH]c(=O)c1[END_SMILES],
property: |LogP|-0.95, |NumHAcceptors|8. Completed molecule: [START_SMILES]

<s> Masked molecule [START_SMILES]O=C(Nc1cc2c(s1)CCC2)c1cc[nH]c(=O)c1[END_SMILES],
property: |LogP|-1.95, |NumHAcceptors|9. Completed molecule: [START_SMILES]
\end{verbatim}

Here, the values of LogP (1.05, 0.05, -0.95, -1.95) and hydrogen bond acceptors (6, 7, 8, 9) were adjusted to guide molecule generation toward the desired properties.

\item Step 3: Candidate molecule generation

The work agent generated the following candidate SMILES molecules:
\begin{itemize}
    \item \texttt{CCn1cc(C)c(C(=O)Nc2sc3c(c2C(N)=O)CCC3)c(=O)c1=O}
    \item \texttt{Cc1cc(C(=O)Nc2sc3c(c2C(N)=O)CCC3)cc(=O)n1C }
    \item \texttt{CCOC(=O)c1c(NC(=O)c2c(C)c(C)n(CC(=O)OC)c(=O)c2C)\linebreak sc2c1[C@@H](C(=O)OC)CC2 }
    \item \texttt{CCOC(=O)c1c(NC(=O)c2c(C)c(C(=O)[O-])n(CC(O)C(O)C(O)CO)c(=O)c2C)\linebreak sc2c1[C@@H](C(=O)[O-])CC2 }
\end{itemize}

\item Step 4: Filtering the optimal molecule

The research agent applied a filter to identify the optimal molecule:
\begin{itemize}
    \item \textbf{Water Solubility Score:} Estimated using molecular weight, TPSA, polarizability, counts of hydroxyl, carboxyl, amino, aromatic rings, halogen, and ether groups.
    \item \textbf{Hydrogen Bond Acceptor (HBA) Score:} Calculated based on counts of hydroxyl, carbonyl, amino, ether, phosphate, carboxylate, nitrile, amide, thioether, and sulfonyl groups.
    \item \textbf{Scoring Weights:} Regression-model-trained weights were applied to individual scores for each property.
    \item \textbf{Ranking:} Candidates were ranked by the sum of weighted scores.
\end{itemize}

\item Step 5: Get final generated molecule results

The molecule with the highest score was:
\[
\texttt{CCOC(=O)c1c(NC(=O)c2c(C)c(C(=O)[O-])n(CC(O)C(O)C(O)CO)c(=O)c2C)sc2c1[C@@H](C(=O)[O-])CC2}
\]
This molecule exhibited a LogP value of -2.83 and an HBA count of 14.0, demonstrating significant improvements in both water solubility and hydrogen bond acceptor capacity.

\end{itemize}

\subsection*{Metric}
We evaluated molecular optimization performance using four key metrics. First, we defined two success metrics: Hit Ratio (Loose), which indicates successful optimization with any positive improvement in target properties, and Hit Ratio (Strict), which requires exceeding specific thresholds (LogP > 1.0, QED > 0.1, tPSA > 10, and changes in hydrogen bond acceptors/donors > 1). Additionally, we assessed structural consistency through the Same Scaffold Ratio, calculated by extracting molecular scaffolds using RDKit and measuring preservation between source and generated molecules. Finally, we evaluated molecular validity through the Valid Ratio, which quantifies the proportion of chemically valid structures among generated molecules.

\subsection*{Experimental Settings}
We conducted our experiments using both LoRA and full parameter fine-tuning approaches on a hardware setup comprising an Intel(R) Xeon(R) Platinum 8468 CPU and NVIDIA A100 80GB GPUs. For models with 125M, 1.3B, and 6.7B parameters, fine-tuning was performed on a single A100 GPU, while the 30B parameter model required 4 A100 GPUs. Each backbone model was fine-tuned on a dataset of 1,000,000 molecules. All datasets used in our experiments are publicly available in our open-source GitHub repository to ensure reproducibility.

For LoRA fine-tuning, we employed the following configuration: rank (r) of 64, scaling parameter of 16, and dropout probability of 0.1 in LoRA layers. The training process utilized mixed precision with FP16 enabled and BF16 disabled. Both training and evaluation batch sizes were set to 188 per GPU device. To optimize memory usage, we enabled gradient checkpointing and set gradient accumulation steps to 1, with maximum gradient norm capped at 0.3 for gradient clipping.
Besides, we compare two training strategies: focus task learning (training separate models for each task) and transfer learning (training all tasks simultaneously). As shown in Extended Data Figure 5, the transfer learning approach achieves comparable or even superior performance compared to focus task learning across multiple metrics including hit ratio, valid ratio, and same scaffold ratio.  This unified training strategy significantly enhances the model's practical utility by eliminating the need for task-specific models. We hypothesize that this effectiveness stems from the inherent correlations between different molecular properties - during transfer learning, the model can learn these inter-property relationships, thereby improving its overall performance. For instance, modifications affecting LogP often impact TPSA simultaneously, and understanding such correlations through joint training helps the model make more informed optimizations.

For optimization, we utilized the Paged AdamW 32-bit optimizer with the following parameters: initial learning rate of 1e-4, weight decay of 0.001 (excluding bias and LayerNorm weights), and a constant learning rate schedule with 0.03 linear warmup ratio. Training was conducted for 1 epoch with a maximum sequence length of 188. We implemented regular model checkpointing every 10,000 steps and logging every 10 steps. For computational efficiency, we disabled sequence length grouping and input packing optimizations.

All prompts used in our experiments are shown in Extended Data Table 4, which provides a comprehensive collection of prompts for both the worker agent and research agent across different optimization tasks. These prompts were carefully designed and validated to ensure consistent performance in molecular optimization scenarios.

The experiments were conducted on a high - performance computing machine equipped with a single Nvidia A100 GPU (80GB) and an Intel(R) Xeon(R) Platinum CPU. For the baseline model, the specific library versions used were as follows: ogb version 1.2.0, megatron-lm version 1.1.5, pyg version 2.0.3, rdkit version 2020.09.1.0, and pytorch version 1.9.1, CUDA version 11.8, Python version 3.9. In our experimental setup, we utilized a distinct set of library versions: transformers version 4.43.0, cuda-NCCL version 2.20.5, numpy version 2.1.2, datasets version 3.0.1, pandas version 2.2.3, CUDA version 11.8, Python version 3.12, and Ubuntu version 22.04, pytorch version 1.9.1. These carefully selected configurations ensured a controlled and consistent environment, which was crucial for making a meaningful and reliable comparison between our results and those of the baseline model.

\subsection*{Data availability}
The datasets utilized in this study are entirely open-source and publicly available to ensure straightforward replication of our findings. For research related to quantum mechanics, physical chemistry, biophysics, and physiology, the datasets can be accessed through PubChem (https://pubchem.ncbi.nlm.nih.gov) \cite{kim2021pubchem} and the Public Access to Neuroactive Anticonvulsant Chemical Evaluations (PANAChE) database (http://panache.ninds.nih.gov).

\subsection*{Code availability}
The code implementation of \model\ is publicly available at \url{https://github.com/jiajunyu1999/LLM4Drug}. This encompasses our methodologies for molecular optimization, including the worker agent and research agent implementations, training procedures, and inference pipelines. 

Furthermore, we are committed to releasing all fine-tuned LLM checkpoints to facilitate reproducibility and enable broader scientific impact. We are also developing an interactive web platform to help scientists utilize \model\ for molecular optimization tasks. The platform will feature comprehensive functionalities including molecular property prediction, optimization guidance, and literature-based knowledge synthesis. Details about accessing the web interface will be provided in supplementary materials. As part of our ongoing commitment to open science, we plan to continuously expand the platform's capabilities in future development phases.

%% file: related.tex
\section*{Related Work}

\subsection*{Large Language Models}
The field of Natural Language Processing (NLP) has undergone transformative advancements with the emergence of large language models (LLMs). The introduction of the Transformer architecture \cite{vaswani2017attention} marked a pivotal milestone, serving as the backbone for numerous state-of-the-art models. Subsequently, LLMs such as BERT \cite{devlin2018bert}, GPT \cite{radford2018improving}, and GPT-2 \cite{radford2019language} demonstrated remarkable capabilities in a variety of linguistic tasks, pushing the boundaries of what NLP systems could achieve. GPT \cite{radford2018improving} revolutionized the field by utilizing generative pre-trained transformers for autoregressive prediction, establishing a potent paradigm for language modeling. Recent groundbreaking advancements, including ChatGPT, GPT-4 \cite{achiam2023gpt}, and LLaMA \cite{dubey2024llama}, have further extended the capabilities of LLMs. These models, trained on vast amounts of text data, exhibit exceptional performance in complex linguistic tasks, showcasing abilities such as contextual reasoning, in-depth understanding of human language, and robust generalization. To fully exploit the potential of pre-trained LLMs, instruction tuning \cite{wei2021finetuned, ouyang2022training,hu2021lora} has emerged as a critical strategy for generating high-quality outputs. By fine-tuning LLMs using instruction-response pairs, this approach aligns model behavior with user-intended outcomes. Open-source models such as Alpaca \cite{taori2023alpaca} and Vicuna \cite{chiang2023vicuna}, which improve upon LLaMA \cite{dubey2024llama}, have demonstrated the utility of instruction tuning in enhancing the adaptability of LLMs to downstream tasks. 

The versatility of LLMs extends beyond traditional NLP tasks, finding applications in diverse domains, including data mining\cite{pan2024integrating}, scientific research\cite{meyer2023chatgpt,zheng2025large,zheng2024large} and molecule design\cite{liu2024conversational,liu2023multi}. This cross-domain adaptability has opened new possibilities for addressing complex problems, particularly in fields like drug discovery, where multi-objective optimization plays a central role.

\subsection*{Molecule Optimization}
Molecular optimization \cite{hoffman2022optimizing,kenakin2003predicting,klebe2015applying} is a critical yet challenging task in drug discovery, involving the design or modification of molecules to achieve multiple objectives such as enhancing biological activity \cite{xie2018selectivity}, improving selectivity \cite{wermuth2004selective}, and minimizing toxicity \cite{narsinghani2014lead}. This process can be broadly categorized into traditional approaches, deep learning-based methods, and recent advancements utilizing large language models (LLMs). In this section, we provide an overview of these approaches, discuss their strengths and limitations, and highlight the motivation behind our proposed system.

Traditional molecular optimization methods rely heavily on domain expertise and heuristic rules \cite{sahinidis2000applications,lipinski1997experimental}, such as Lipinski's Rule of Five \cite{lipinski1997experimental}, to guide the modification of molecular structures. These approaches typically involve iterative manual adjustments followed by extensive experimental validation, making them time-consuming and resource-intensive. While effective in certain cases, these methods are limited by their dependence on human intuition, resulting in a subjective and difficult-to-scale process. Additionally, traditional approaches often struggle to balance conflicting objectives, such as maximizing efficacy while minimizing toxicity, due to the lack of a systematic framework.

The advent of deep learning has introduced generative models, including variational autoencoders (VAEs) \cite{kingma2013auto}, generative adversarial networks (GANs) \cite{goodfellow2014generative}, and reinforcement learning (RL) algorithms \cite{arulkumaran2017deep,sutton2018reinforcement}, as powerful tools for molecular optimization. These methods automate molecule generation by learning patterns from molecular datasets and optimizing structures to achieve desired properties. However, deep learning models often operate in a purely data-driven manner, making it challenging to incorporate prior knowledge or leverage insights from scientific literature. Furthermore, these approaches frequently lack interpretability, limiting their acceptance in applications that require a clear understanding of the optimization process.

Recent advancements in large language models (LLMs) have opened new possibilities for molecular optimization. Models such as LigGPT \cite{bagalliggpt} have demonstrated the ability to generate molecules with specific scaffolds and properties by leveraging the expressiveness of transformer architectures. Controlled optimization frameworks like Transformer-R \cite{he2021transformer} incorporate chemical property constraints to ensure that optimized molecules retain desired features. Similarly, approaches such as Modof-pipe \cite{chen2021deep} focus on preserving molecular scaffolds while allowing controlled structural modifications, thereby ensuring chemical validity and relevance to specific applications. MoleculeSTM \cite{liu2023multi} employs multimodal learning to link molecular structures with textual descriptions, enhancing the alignment between optimization objectives and model outputs. Prompt-MolOpt \cite{wu2024leveraging} improves the optimization process by allowing user-defined fragments to guide structural transformations. Additionally, conversational frameworks like ChatDrug \cite{liu2024conversational} integrate LLMs with domain-specific feedback to iteratively refine optimization strategies.

Despite these advancements, LLM-based methods for molecular optimization face significant challenges. They typically require large molecular pair datasets, which are difficult and costly to obtain, limiting the diversity and success rate of generated molecules. Moreover, ensuring scaffold consistency—a critical factor in molecular design—remains a major hurdle, as these models often struggle to precisely control molecular modifications. This lack of control compromises chemical validity and alignment with specific optimization goals.

%% file: extented_data.tex
\section*{Extended Data}

\begin{figure}[h!]
\caption*{\textbf{Extended Data Table 1 | Features Used for Estimating LogP, QED, TPSA, HBA, and HBD Scores.} A comprehensive summary of the molecular features utilized in estimating the LogP, QED, TPSA, HBA, and HBD scores, based on various chemical properties and functional groups present in the molecules.}
\renewcommand{\arraystretch}{1.5}
\begin{tabular}{|>{\centering\arraybackslash}m{6cm}|>{\centering\arraybackslash}m{11cm}|}
\hline
\textbf{Score} & \textbf{Features Used} \\
\hline
LogP & Molecular Weight (MW), TPSA, Hydrogen Bond Donors (HBD), Hydrogen Bond Acceptors (HBA), Polarizability, Hydroxyl Groups, Carboxyl Groups, Amino Groups, Aromatic Rings, Halogens, Ether Groups \\
\hline
QED & Molecular Weight (MW), LogP, TPSA, Hydrogen Bond Donors (HBD), Hydrogen Bond Acceptors (HBA), Aromatic Rings, Rotatable Bonds, Charge, Stereochemistry, Solubility, Synthetic Accessibility (SA) \\
\hline
TPSA & Hydroxyl Groups, Carboxyl Groups, Amino Groups, Carbonyl Groups, Sulfonamide Groups, Ether Groups, Thiol Groups, Phosphate Groups, Amide Groups, Nitro Groups \\
\hline
HBA & Hydroxyl Groups, Carbonyl Groups, Amino Groups, Ether Groups, Phosphate Groups, Carboxylate Groups, Nitrile Groups, Amide Groups, Thioether Groups, Sulfonyl Groups \\
\hline
HBD & Carbonyl Groups, Nitro Groups, Ether Groups, Amide Groups, Aromatic Nitro Compounds, Sulfonyl Groups, Phosphoryl Groups, Hydroxyl Groups, Imidazole Rings, Carboxylate Ions \\
\hline
\end{tabular}
\label{ED_tab5}
\end{figure}

\begin{table}[h!]
\caption*{\textbf{Extended Data Table 2 | Task Descriptions for Single and Multi-Property Optimization.} This table presents the complete set of optimization tasks evaluated in this study, including both single property modifications and combined property adjustments.}
\begin{tabular}{lll}
\toprule
\textbf{Task ID} & \textbf{Description} & \textbf{Strict Threshold} \\
\midrule
\multicolumn{3}{c}{\textit{Single Property Optimization}} \\
\midrule
101 & Increase water solubility (Decrease LogP) & LogP: 0.5 \\
102 & Decrease water solubility (Increase LogP) & LogP: 0.5 \\
103 & Increase drug-likeness (Increase QED) & QED: 0.1 \\
104 & Decrease drug-likeness (Decrease QED) & QED: 0.1 \\
105 & Increase permeability (Decrease TPSA) & TPSA: 10 \\
106 & Decrease permeability (Increase TPSA) & TPSA: 10 \\
107 & Increase hydrogen bond acceptors (\# HBA) & \# HBA: 1 \\
108 & Increase hydrogen bond donors (\# HBD) & \# HBD: 1 \\
\midrule
\multicolumn{3}{c}{\textit{Multiple Properties Optimization}} \\
\midrule
201 & Increase water solubility (Decrease LogP) and hydrogen bond acceptors (\# HBA) & LogP: 0.5, \# HBA: 1 \\
202 & Decrease water solubility (Increase LogP) and hydrogen bond acceptors (\# HBA) & LogP: 0.5, \# HBA: 1 \\
203 & Increase water solubility (Decrease LogP) and hydrogen bond donors (\# HBD) & LogP: 0.5, \# HBD: 1 \\
204 & Decrease water solubility (Increase LogP) and hydrogen bond donors (\# HBD) & LogP: 0.5, \# HBD: 1 \\
205 & Increase water solubility (Decrease LogP) and permeability (Decrease TPSA) & LogP: 0.5, TPSA: 10 \\
206 & Increase water solubility (Decrease LogP) and decrease permeability (Increase TPSA) & LogP: 0.5, TPSA: 10 \\
\bottomrule
\end{tabular}
\label{ED_tab1}
\end{table}


\begin{figure}[h!]
\caption*{\textbf{Extended Data Table 3 | Task Performance Metrics for Single and Multi-Property Optimization on Unseen Smiles.} This table presents comprehensive performance metrics across different optimization tasks, showing the model's effectiveness in both single and multi-property scenarios. The metrics include hit ratios under loose and strict criteria, molecule validity rates, and scaffold similarity measures.}
\begin{tabular}{lcccc}
\toprule
\textbf{Task ID} & \textbf{Hit Ratio (Loose)} & \textbf{Hit Ratio (Strict)} & \textbf{Valid Ratio} & \textbf{Same Scaffold Ratio} \\
\midrule
\multicolumn{5}{c}{\textit{Single Property Optimization}} \\
\midrule
101 & 91.37 & 82.01 & 93.71 & 89.45 \\
102 & 94.51 & 88.28 & 90.11 & 92.77 \\
103 & 69.75 & 34.09 & 99.55 & 85.33 \\
104 & 82.38 & 58.81 & 90.56 & 94.79 \\
105 & 76.68 & 60.34 & 93.48 & 96.88 \\
106 & 96.88 & 91.59 & 93.48 & 93.03 \\
107 & 99.28 & 95.91 & 93.48 & 93.75 \\
108 & 99.04 & 96.15 & 93.48 & 91.83 \\
\midrule
\multicolumn{5}{c}{\textit{Multiple Properties Optimization}} \\
\midrule
201 & 90.63 & 86.54 & 93.48 & 94.47 \\
202 & 58.35 & 34.19 & 87.42 & 92.80 \\
203 & 95.81 & 89.66 & 91.24 & 92.86 \\
204 & 51.87 & 28.68 & 90.11 & 90.27 \\
205 & 29.26 & 11.27 & 93.71 & 94.72 \\
206 & 83.97 & 66.03 & 93.93 & 94.26 \\
\bottomrule
\end{tabular}
\label{ED_tab1}
\end{figure}

\begin{figure}[h!]
\caption*{\textbf{Extended Data Table 4 | Prompt Templates Used in MultiMol.} A detailed overview of prompt templates employed for model training and inference across different tasks, including both single-molecule optimization and molecule-protein interaction scenarios.}
\renewcommand{\arraystretch}{1.5}
\begin{tabular}{|>{\centering\arraybackslash}m{6cm}|>{\centering\arraybackslash}m{11cm}|}
\hline
\textbf{Task Type} & \textbf{Template Format} \\
\hline
\textbf{MultiMol Single-Molecule Training} & 
\texttt{<s> Masked SMILES: [START\_SMILES] \{ \} [END\_SMILES], properties: \{ \}. Completed SMILES: [START\_SMILES] \{ \} [END\_SMILES]. </s>} \\
\hline
\textbf{Alternative LLM Training} & 
\texttt{Please recover this masked molecule \{ \}, the desired properties are \{ \}. The recovered molecule is \{ \}} \\
\hline
\textbf{MultiMol Protein-Ligand Training} & 
\texttt{<s> Masked SMILES: [START\_SMILES] \{ \} [END\_SMILES], property: \{ \}, the affinity between the molecule and the protein [START\_AMINO] \{ \} [END\_AMINO] is \{ \}. Completed SMILES: [START\_SMILES] \{ \} [END\_SMILES]. </s>} \\
\hline
\textbf{MultiMol Single-Molecule Inference} & 
\texttt{<s> Masked molecule [START\_SMILES] \{ \}[END\_SMILES], property: \{ \}. Completed molecule: [START\_SMILES]} \\
\hline
\textbf{Alternative LLM Inference} & 
\texttt{Please recover this masked molecule \{ \}, the desired properties are \{ \}. The recovered molecule is} \\
\hline
\textbf{MultiMol Protein-Ligand Inference} & 
\texttt{<s> Masked SMILES: [START\_SMILES] \{ \} [END\_SMILES], property: \{ \}, the affinity between the molecule and the protein [START\_AMINO] \{ \} [END\_AMINO] is \{ \}. Completed SMILES: [START\_SMILES] \{ \} [END\_SMILES]} \\
\hline
\end{tabular}
\label{ED_tab3}
\end{figure}

\begin{table}[h!]
\centering
\caption*{\textbf{Extended Data Table 5 | $\Delta$ for LogP, QED, TPSA, \# HBA, and \# HBD.} 
This table presents the complete set of incremental $\Delta$ values for each property analyzed in this study, providing stepwise changes applied during property optimization.}
\renewcommand{\arraystretch}{1.2} 
\begin{tabular}{>{\bfseries}c>{\centering}p{1.5cm}>{\centering}p{1.5cm}>{\centering}p{1.5cm}>{\centering}p{1.5cm}>{\centering\arraybackslash}p{1.5cm}}
\toprule
\textbf{Index} & \textbf{LogP} & \textbf{QED} & \textbf{TPSA} & \textbf{\# HBA} & \textbf{\# HBD} \\
\midrule
1  & 0.5  & 0.05 & 5  & 1  & 1  \\
2  & 1.0  & 0.10 & 10 & 2  & 2  \\
3  & 1.5  & 0.15 & 15 & 3  & 3  \\
4  & 2.0  & 0.20 & 20 & 4  & 4  \\
5  & 2.5  & 0.25 & 25 & 5  & 5  \\
6  & 3.0  & 0.30 & 30 & 6  & 6  \\
7  & 3.5  & 0.35 & 35 & 7  & 7  \\
8  & 4.0  & 0.40 & 40 & 8  & 8  \\
9  & 4.5  & 0.45 & 45 & 9  & 9  \\
10 & 5.0  & 0.50 & 50 & 10 & 10 \\
11 & 5.5  & 0.55 & 55 & 11 & 11 \\
12 & 6.0  & 0.60 & 60 & 12 & 12 \\
13 & 6.5  & 0.65 & 65 & 13 & 13 \\
14 & 7.0  & 0.70 & 70 & 14 & 14 \\
15 & 7.5  & 0.75 & 75 & 15 & 15 \\
\bottomrule
\end{tabular}
\label{delta_values}
\end{table}

\begin{figure}[h!]
\caption*{\textbf{Extended Data Figure 1 | MultiMol Recovery Performance.} Density plots illustrating the ability of MultiMol to recover the original molecules using the prompt $(Source\ Property + Scaffold) \rightarrow Source\ Molecule$. Each subfigure shows the distribution of specific molecular properties (MolLogP, QED, TPSA, NumHAcceptors, and NumHDonors) for both the source and recovered molecules, with the mean values indicated by dashed lines.}
\includegraphics[width=1.0\linewidth]{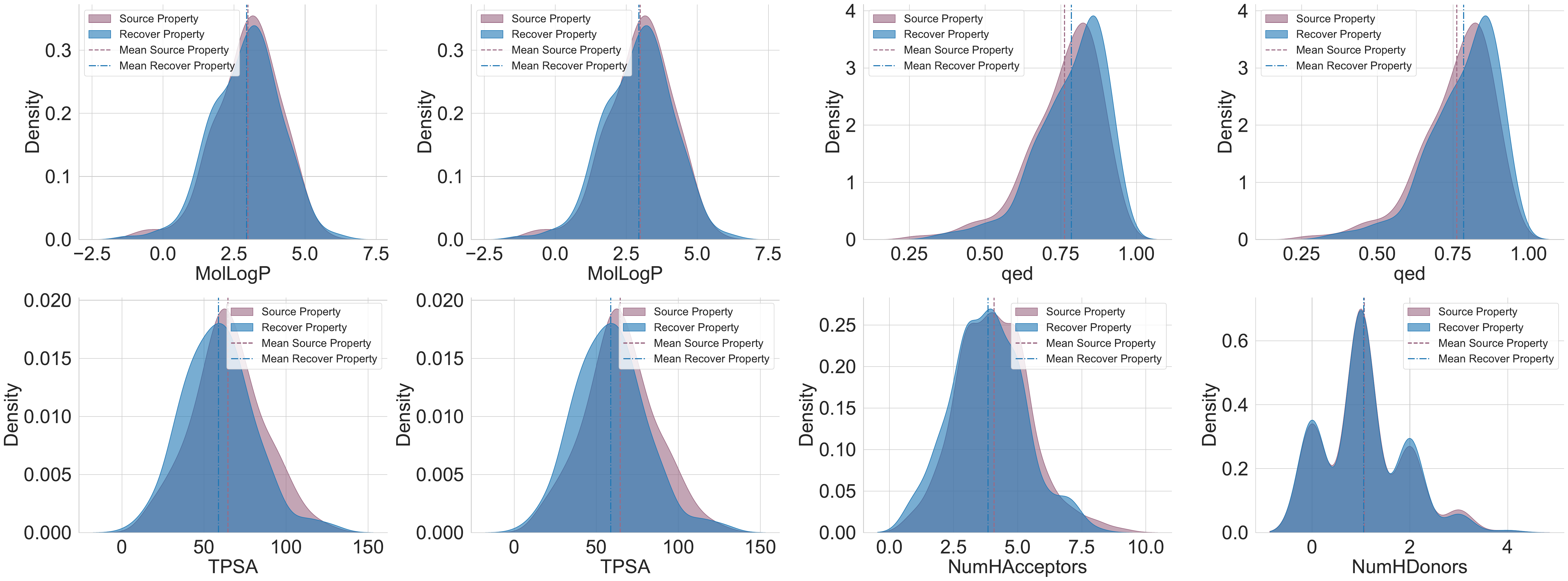}
\label{ED_fig1}
\end{figure}

\begin{figure}[h!]
\caption*{\textbf{Extended Data Figure 2 | Impact of candidate molecule pool size on optimization metrics.} Analysis of how varying the candidate pool size (1, 2, 4, 8) influences optimization performance for both single-property (top) and multi-property (bottom) tasks. The distributions of key molecular properties (MolLogP, QED, TPSA, NumHAcceptors, and NumHDonors) demonstrate how expanding the search space affects the model's ability to identify optimal molecules.}
\includegraphics[width=1.0\linewidth]{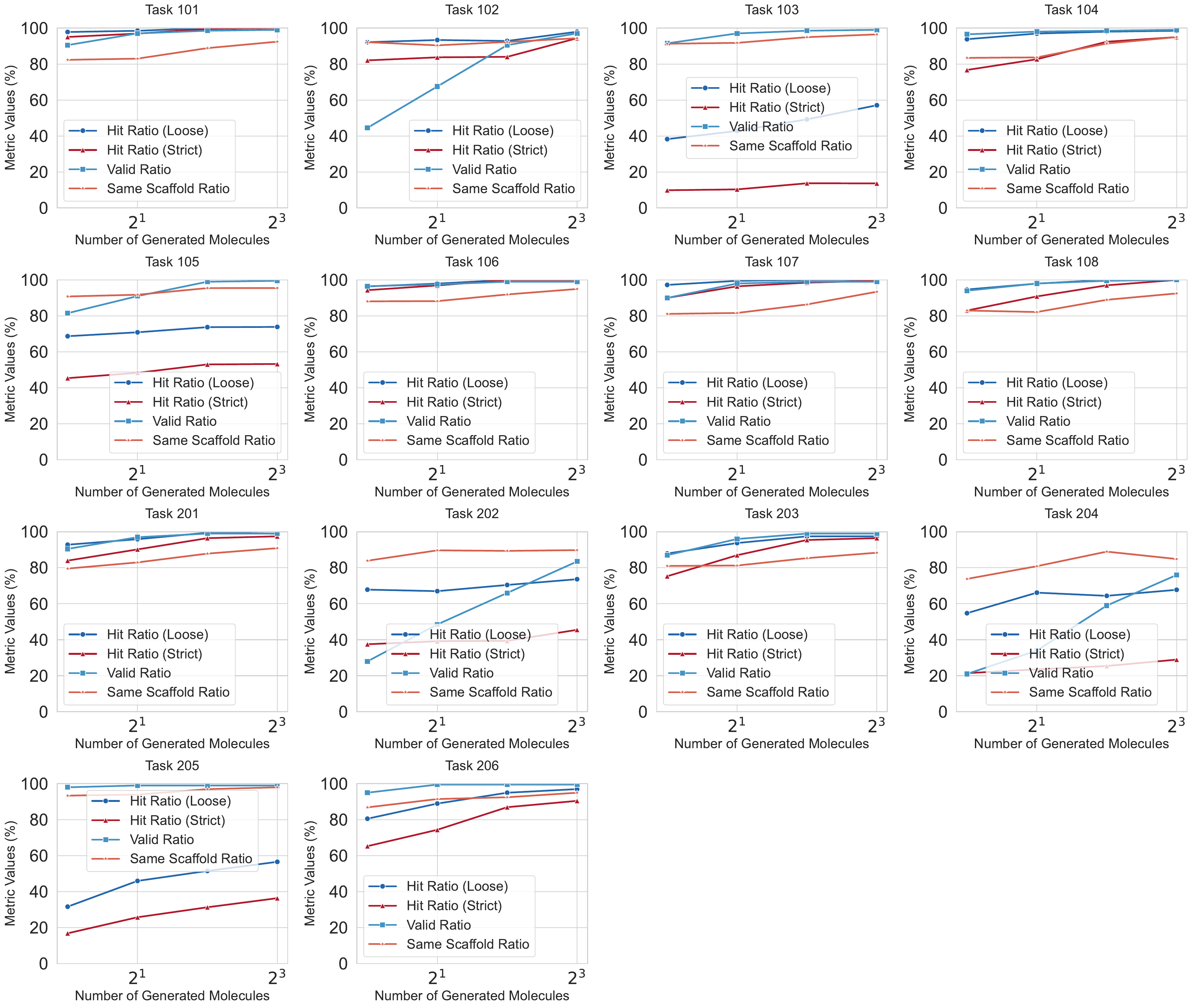}
\label{ED_fig2}
\end{figure}

\begin{figure}[h!]
\caption*{\textbf{Extended Data Figure 3 | Performance Comparison Across Different Backbones.} Comparison of performance metrics between Qwen2.5-7b, Llama3.1-8b, Galactica-6.7b models across different optimization tasks. Each subplot shows a specific metric (Hit Ratio (Loose), Hit Ratio (Strict), Valid Ratio, and Same Scaffold Ratio) for both single and multi-property optimization tasks, with task IDs on the x-axis and corresponding performance values on the y-axis.}
\includegraphics[width=1.0\linewidth]{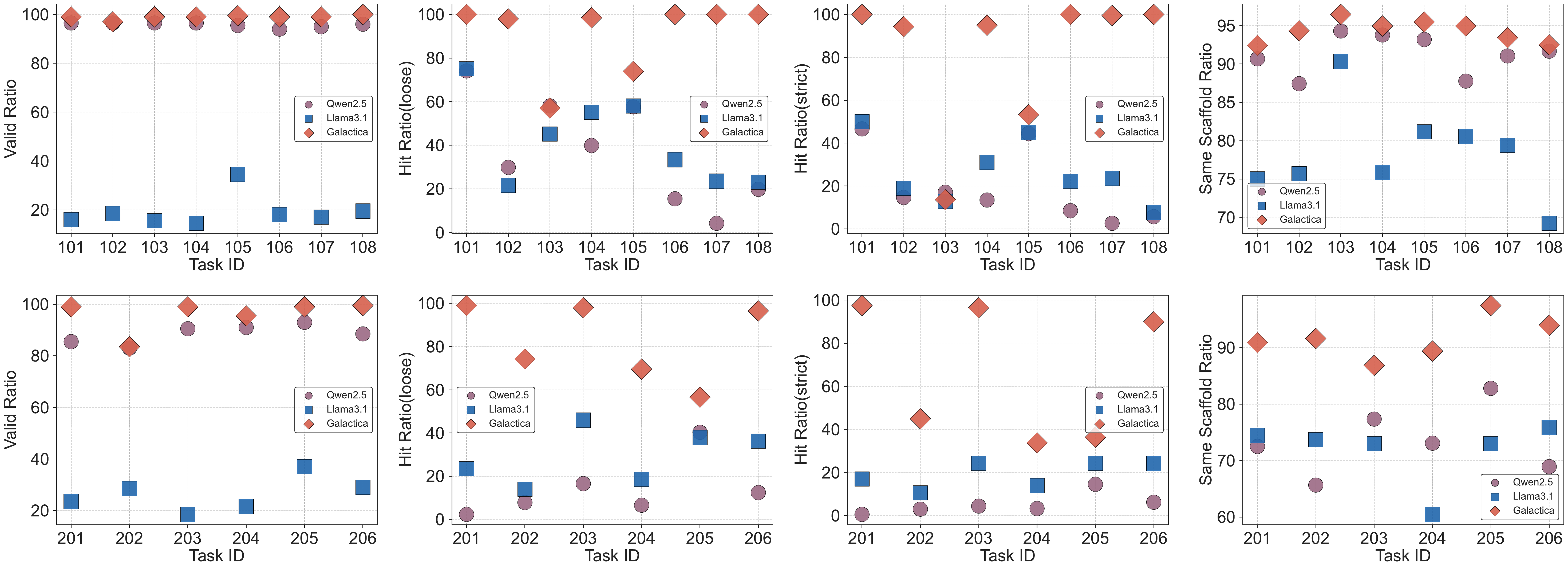}
\label{ED_fig2}
\end{figure}

\begin{figure}[h!]
\caption*{\textbf{Extended Data Figure 4 | Score Differences Between Generated and Source Molecules.} Box plots showing the score differences ($\Delta$Score = Generated Score - Source Score) across different functional groups for each single-property optimization task (Tasks 101-108). The plots are divided into positive (blue shaded) and negative (red shaded) regions, indicating whether the generated molecules achieved higher or lower scores than the source molecules. Each box shows the quartiles of the score differences, with whiskers extending to show the rest of the distribution. Individual points represent outliers.}
\includegraphics[width=1.0\linewidth]{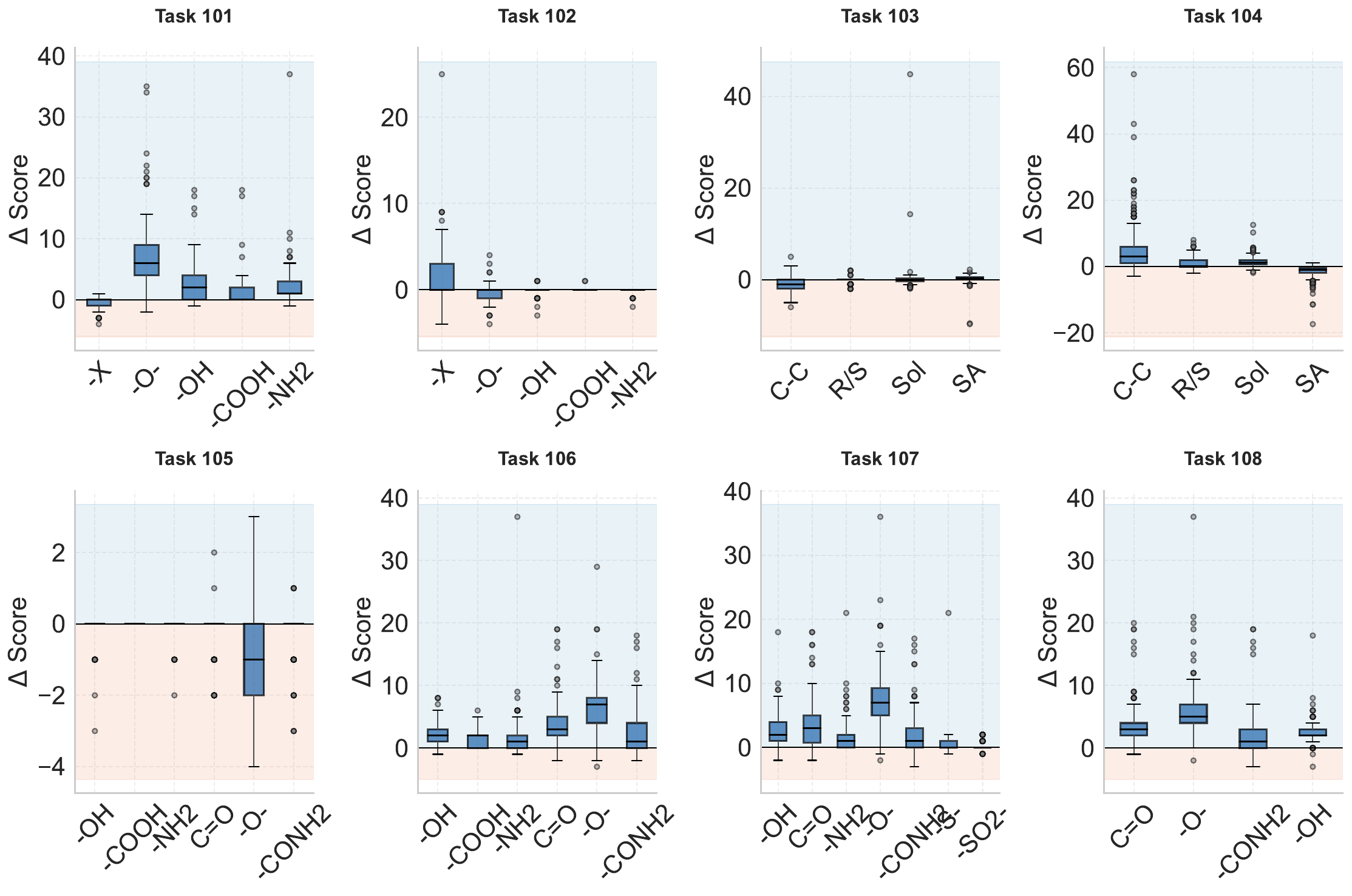}
\label{ED_fig3}
\end{figure}

\begin{figure}[h!]
\caption*{\textbf{Extended Data Figure 5 | Performance comparison between joint training and task-specific fine-tuning.} The figure compares two training strategies: joint training using all tasks simultaneously (Transfer Learning) versus individual task-specific (Focus Task Learning) fine-tuning, using Galactica-6.7B model with LoRA fine-tuning on 1M samples. The performance metrics shown include success rate, validity rate, and property improvement scores across different optimization tasks.}
\includegraphics[width=1\linewidth]{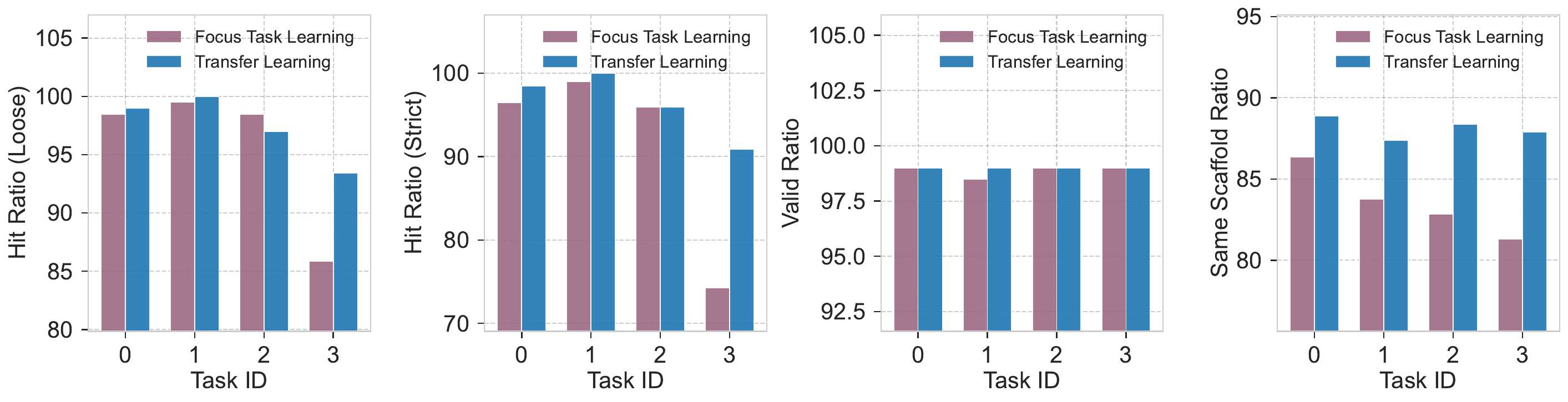}
\label{ED_fig4}
\end{figure}